# Experimental Tests of Spirituality

Abraham Loeb

We currently harness technologies that could shed new light on old philosophical questions, such as whether our mind entails anything beyond our body or whether our moral values reflect universal truth.

———

By Abraham Loeb on June 4, 2018

For thousands of years philosophers pondered on deep questions, such as whether spirituality goes beyond the makeup of our body and allows free will and whether moral values reflect universal truth similarly to the laws of physics. Could we put their proposed answers to the test of modern science? Could we use experimental data to advance our knowledge on these questions?

Historically, psychological or anthropological studies of people addressed the possible illusion of free will (for example, by Dan Wegner[1]) or the multi-cultural variations in moral values[2]. But observing a complex system like a human without having its blueprint or without the ability to dissect its constituents and reconstruct them from scratch, is far more challenging than figuring out how the hardware of an iPhone works by just interacting with it.

There are at least three new experimental frontiers that offer prospects for advances:

1. *Artificial Intelligence and Machine Learning*: the latest development of self-learning computer algorithms allows us to envision a future computer system that learns from experience and educates itself about the world like a human. Such a system could interact with the physical world around it through sensors of vision, sound, taste, smell and touch, like a human. It might even reproduce itself using a 3D printer, in analogy to child birth. If such a "pseudo-human" system were to exhibit the qualitative features of the phenomenon of "free will" in a fashion that is completely indistinguishable from a human, then one could conclude that free will is an emergent phenomenon which results from putting together well-understood components. Restructuring those building blocks might lead to predictable changes in the decisions of the system. If so, one might be able to demonstrate that the will of the system can be manipulated through artificial design of its constituents. The will exhibited by the system will not be "free" at a fundamental level if it can be manipulated and controlled artificially.

2. *Laboratory experiments on the Origins of Life*: several teams are currently engaged in experiments which are aimed to demonstrate that life can be produced synthetically in a controlled laboratory environment. If any of these teams is successful at producing *artificial life* from a soup of chemicals, that would constitute a conceptual breakthrough. Discovering a chemical recipe for life would qualitatively resemble finding the recipe for a cake, which has no spiritual component beyond its material ingredients. Such understanding could be combined with the technology of genome editing to manipulate the higher level blueprint of living systems.

3. The *Search for Extraterrestrial Intelligence*: the recent survey by the *Kepler* satellite revealed that about a quarter of all stars in the Milky Way galaxy host habitable Earth-size planets, where the chemistry of "life-as-we-know-it" can develop. It would be presumptuous of us to contemplate that we are alone. Out of modesty, we should assume that other civilizations exist out there. Discovering them will enable us to test whether our moral values reflect universal truth through a census of the cultures we find on exoplanets.

There is no doubt that advances on any of these fronts would reshape our philosophical perspective on the above-mentioned mind-body or universality-of-ethics questions. Science has done it before by shaping our culture through the development of the technologies that enabled nuclear bombs or a trip to the Moon. There is no doubt that Socrates, Plato or Aristotle would have been fascinated by these new opportunities to experiment with their ideas. We are fortunate to live at an exciting time where philosophical conjectures could become testable scientific hypotheses on their path to serving as robust pillars for our understanding of reality.

## References

1. Wegner, D. M., "The Illusion of Conscious Will", MIT Press (2002).
2. Gowans, C. "Moral Relativism", *The Stanford Encyclopedia of Philosophy, edited by Edward N. Zalta* (2017); https://plato.stanford.edu/entries/moral-relativism/

### ABOUT THE AUTHOR

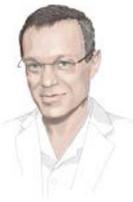

**Abraham Loeb**

Abraham Loeb is chair of the astronomy department at Harvard University, founding director of Harvard's Black Hole Initiative and director of the Institute for Theory and Computation at the Harvard-Smithsonian Center for Astrophysics. He chairs the Board on Physics and Astronomy of the National Academies and the advisory board for the Breakthrough Starshot project.

Credit: Nick Higgins